\documentclass[aps,prl,
longbibliography,
preprintnumbers,
reprint
]{revtex4-2}
\pdfoutput=1

\usepackage{amsfonts}
\usepackage{amsmath}
\usepackage{amssymb}
\usepackage{amsthm}
\usepackage[mathscr]{eucal}
\usepackage{bbold}
\usepackage{braket}
\usepackage{color}
\usepackage{dsfont}
\usepackage{framed}
\usepackage{marginnote}
\usepackage{mathtools}
\usepackage{physics}
\usepackage[normalem]{ulem}
\usepackage{stackrel}
\usepackage{tensor}
\usepackage{tikz-cd}
\usepackage{ulem}
\usepackage{subcaption}
\usepackage[export]{adjustbox}

\usepackage{hyperref}

\usepackage{cleveref}
\crefname{appendix}{App.}{Apps.}

\makeatletter\def\@fpheader{~}\makeatother
\hypersetup{
	colorlinks,
	linkcolor={blue!80!black},
	citecolor={blue!80!black},
	urlcolor={blue!80!black}
}

\newcommand{\smalltext}[1]{\text{\fontsize{1pt}{0}\selectfont#1}}
\newcommand{\reprel}[2]{\stackrel{\scalebox{0.5}{$#1$}}{#2}}

\DeclareMathOperator{\Aut}{Aut}

\definecolor{SHCcolor}{rgb}{0.9,0.1,0.1}

\newcommand{\ignore}[1]{}
\definecolor{NVMcolor}{rgb}{0.5,0.7,0.1}

\definecolor{WWWcolor}{rgb}{.1,..1,.9}

\begin{document}

\preprint{MIT-CTP/5814}

\title{A single geometry from an all-genus expansion in quantum gravity}

\author{Sergio Hern\'andez-Cuenca} \email{sergiohc@mit.edu}
\author{Nico Valdes-Meller} \email{nvaldes@mit.edu}
\affiliation{Center for Theoretical Physics, Massachusetts Institute of Technology \\ Cambridge, MA 02139 USA
}%

\author{Wayne Wei-en Weng} \email{www62@cornell.edu}

\affiliation{
 Department of Physics,\\Cornell University \\ Ithaca, NY 14850 USA
}%

\begin{abstract}
We report on an instance in quantum gravity where a topological expansion resums into an effective description on a single geometry. The original theory whose gravitational path integral we study is JT quantum gravity with one asymptotic boundary at nonperturbatively low temperatures. The effective theory we derive is a deformation of JT gravity by a highly quantum and nonlocal interaction for the dilaton, evaluated only on a disk topology. This emergent description addresses a strongly quantum gravitational regime where all genera contribute at the same order, successfully capturing the doubly nonperturbative physics of the original theory.
\end{abstract}

\maketitle

{\bf Introduction---} Quantum gravity can be formulated as a gravitational path integral (GPI) over geometries consistent with some prescribed conditions.
The most general incarnation of the GPI involves integrating over all possible metrics a given topological manifold can be endowed with, and also summing over topologies. When no single topology dominates over the rest, a traditional semiclassical treatment around a leading saddle would seem to be unavailable. 

This raises the question: could a sum over topologies give rise to an effective description involving a path integral over a single topology? 
Our main motivation to address this issue is to understand how one may recover an effective spacetime description in regimes involving highly quantum gravitational superpositions of geometries. This is particularly crucial for exploring quantum gravity in higher dimensions, where in general the GPI cannot be computed beyond semiclassics.

Various authors in the 80's entertained this idea, and studied potential physical consequences for our universe of integrating out topological fluctuations \cite{Hawking:1982dj, Coleman:1988cy, Giddings:1988cx}. 
These fluctuations characteristically induce nonlocal interactions among quantum fields mediated by wormholes, which result in an inherent randomness in the couplings of any local theory for our universe \cite{Giddings:2020yes,Balasubramanian:2020lux,Valdes-Meller:2021com,Hernandez-Cuenca:2024pey}. Recently, these ideas have resurfaced in the context of low-dimensional quantum gravity models and their interpretation as ensembles of theories \cite{Saad:2019lba,Stanford:2019vob,Marolf:2020xie,Belin:2020hea,Maloney:2020nni,Afkhami-Jeddi:2020ezh,Cotler:2020ugk,Collier:2021rsn,Chandra:2022bqq}.

Another setting where this question appears is in string theory, which in perturbation theory involves a worldsheet genus expansion. Strikingly, at strong coupling where higher genus surfaces dominate and the expansion breaks down, it is sometimes the case that an alternative description emerges with a new perturbative expansion around a single leading topology.
The paradigmatic example of this is the S-duality between a fundamental string and a D1-brane in type IIB string theory \cite{Polchinski:1998rr, Green:2005ba}. Another example pertinent to our discussion where D-branes must replace a stringy description is \cite{Drukker:2000rr,Drukker:2005kx}.

The two scenarios above illustrate opposite limits of the phenomenon of a topological expansion admitting an effective description in terms of a single geometry. The first case occurs at ``weak'' coupling, where higher topologies are suppressed, while the second is at ``strong'' coupling, where they are enhanced. In this work we address an expansion at ``intermediate'' coupling, where all topologies actually contribute at the same order.

We study the GPI in Jackiw-Teitelboim (JT) gravity \cite{Jackiw:1984je, Teitelboim:1983ux} at nonperturbatively low temperatures. We use this theory because its topological expansion is tractable \cite{Saad:2019lba}, and tune it to low temperatures to amplify topological fluctuations. At the temperatures we consider there is neither genus suppression nor enhancement; every genus must be included. Our main result is that, even in this most quantum regime of the GPI, an effective description does arise involving a single geometry. Interestingly, our findings at this intermediate coupling exhibit both the nonlocal features of weak coupling and the extra-boundary aspect of D-branes of strong coupling.

Studies of JT gravity at such low temperatures with different motivations include \cite{Okuyama:2019xbv,Okuyama:2020ncd,Engelhardt:2020qpv,Johnson:2020mwi,Johnson:2021rsh,Chandrasekaran:2022asa,Hernandez-Cuenca:2024icn}.
A line of work related in spirit but differing in detail concerns the spectral form factor \cite{Blommaert:2022lbh,Saad:2022kfe,Okuyama:2023pio}.
In particular, the plateau at nonperturbatively late times is found to involve a full genus sum, albeit a different one from ours with \`a priori no analogous effective geometric description.

{\bf Topological resummation---}The observable we focus on is the partition function $\langle Z(\beta)\rangle $ at inverse temperature $\beta$ computed in JT gravity. The GPI for this object is naturally organized by genus into
\begin{align}
\label{eq:topoexp}
    \langle Z(\beta)\rangle = \sum_{g=0}^\infty e^{-(2g-1)S_0} Z_g(\beta),
\end{align}
an asymptotic series at large $S_0$ with the expansion parameter $e^{-S_0}\ll 1$. However, $\beta^{3/2} Z_g(\beta)$ is a polynomial in $\beta$ of degree $3g$, which implies that when $\beta \sim e^{2S_0/3}$ the topological expansion in \cref{eq:topoexp} no longer makes sense. A better organization of the GPI in this regime requires dissecting the computation of $Z_g(\beta)$.

The genus-zero contribution is special and given by
\begin{align}
\label{eq:zerog}
    Z_0(\beta) = \frac{e^{2\pi^2/\beta}}{\sqrt{2\pi} \beta^{3/2}}.
\end{align}
For $g\geq 1$, $Z_g$ involves the Weil-Petersson (WP) volume $V_{g,1}(b)$ of the moduli space of genus $g$ hyperbolic Riemann surfaces with a geodesic boundary of length $b$ \cite{Saad:2019lba},
\begin{align}
\label{eq:zgb}
    Z_g(\beta) = \int_0^\infty b \, db \,  Z_{\smalltext{T}}(\beta,b) \, V_{g,1}(b) ,
\end{align}
where $Z_{\smalltext{T}}(\beta,b)$ is the JT path integral on a trumpet geometry between asymptotic and geodesic boundaries,
\begin{align}
\label{eq:tpet}
    Z_{\smalltext{T}}(\beta,b) = \frac{e^{-b^2/2\beta}}{\sqrt{2\pi \beta}}.
\end{align}
As Mirzakhani showed \cite{Mirzakhani:2006eta}, for general $g$ and geodesic boundary number $n$, the WP volumes read
\begin{equation}
\begin{aligned}
\label{eq:mirz-vol}
    V_{g,n}(b_1,\dots,b_n) &= \sum_{0\leq|{\alpha}|\leq 3g-3+n} C_g({\alpha}) \, b_1^{2\alpha_1} \cdots b_n^{2\alpha_n},\\
    C_g({\alpha}) &= \frac{\expval{\omega^{3g-3+n-|{\alpha}|}\psi_1^{\alpha_1} \cdots \psi_n^{\alpha_n}}_{g,n}}{2^{|\alpha|} {\alpha}!(3g-3+n-|{\alpha}|)!} ,
\end{aligned}
\end{equation}
where the sum is over ${\alpha}\equiv\{\alpha_i\}_{i=1}^n$, $|{\alpha}| = \sum_{i} \alpha_i$,
and ${\alpha}!=\prod_i \alpha_i!$. Here $\expval{\,\cdot\,}_{g,n}$ denotes an integral over moduli space $\overline{\mathcal{M}}_{g,n}$, and defines intersection numbers of first Chern classes $\psi_i$ of the line bundle associated to each boundary, with $\omega$ the WP symplectic form (see e.g. \cite{Do,Mirzakhani:2006eta}).
Using \cref{eq:mirz-vol} in \cref{eq:zgb} gives the anticipated polynomial,
\begin{align}
\label{eq:zgl}
    Z_g(\beta) = \frac{\beta^{3g}}{{\sqrt{2\pi} \beta^{3/2}}}\sum_{\ell=0}^{3g-2} \frac{\beta^{-\ell}}{\ell!} \expval{\omega^{\ell}\psi^{3g-2-\ell}}_{g,1}.
\end{align}
which allows to express $\expval{Z(\beta)}$ in \cref{eq:topoexp} as a double series first in $e^{-S_0}$, then in $1/\beta$. Crucially, scaling $\beta \sim e^{2S_0/3}$ at large $S_0$, the organization that naturally results is first in $1/\beta$, then in $e^{-S_0}$. Explicitly,
\begin{equation}
\begin{aligned}
\label{eq:lexp}
    \expval{ Z(\beta)} &= \frac{1}{\sqrt{2\pi x}} \sum_{\ell=0}^\infty \frac{\beta^{-\ell}}{\ell!} \, F_\ell(x), \quad x\equiv \beta^3 e^{-2S_0}, \\
    F_\ell(x) &\equiv \, (2\pi^2)^\ell + \!\!\!\sum_{g=\lceil (\ell+2)/3\rceil}^\infty \!\!\! x^{g} \, \expval{\omega^{\ell}\psi^{3g-2-\ell}}_{g,1},
\end{aligned}
\end{equation}
with $x\sim O(1)$. Remarkably, the genus sum that defines each $F_\ell$ is automatically convergent. In this scaling limit, the leading term is $\ell=0$, which involves well-known Airy-limit topological gravity intersection numbers,
\begin{align}
\label{eq:g-int-number}
    \expval{\psi^{3g-2}}_{g,1} = \frac{1}{24^g g!}.
\end{align}
Summing over genus, $\expval{Z(\beta)}$ is to leading order given by
\begin{align}
\label{eq:lzero}
    \expval{Z(\beta)} \reprel{\ell=0}{=} \frac{e^{x/24}}{\sqrt{2\pi x}} = \frac{e^{S_0}}{\sqrt{2\pi} \beta^{3/2}} \exp(\frac{\beta^3e^{-2S_0}}{24}). 
\end{align}
Higher-$\ell$ corrections to this result were found in \cite{Okuyama:2019xbv, Okuyama:2020ncd}, and are addressed after the next section. 
Since the standard expansion in \cref{eq:topoexp} admits a semiclassical evaluation of its leading $g=0$ term in \cref{eq:zerog}, a similar treatment would be desirable for the $\ell=0$ in \cref{eq:lzero}. However, in this large-$\beta$ regime, each $F_\ell$ as written in \cref{eq:lexp} involves a genus sum where every term is of the same order. As a very quantum gravitational superposition of all possible topologies, an effective realization of even just $F_0$ on a single geometry is not readily available with this description.

\begin{figure}[ht!]
\begin{subfigure}{.25\textwidth}
  \centering
  \includegraphics[width=0.65\textwidth]{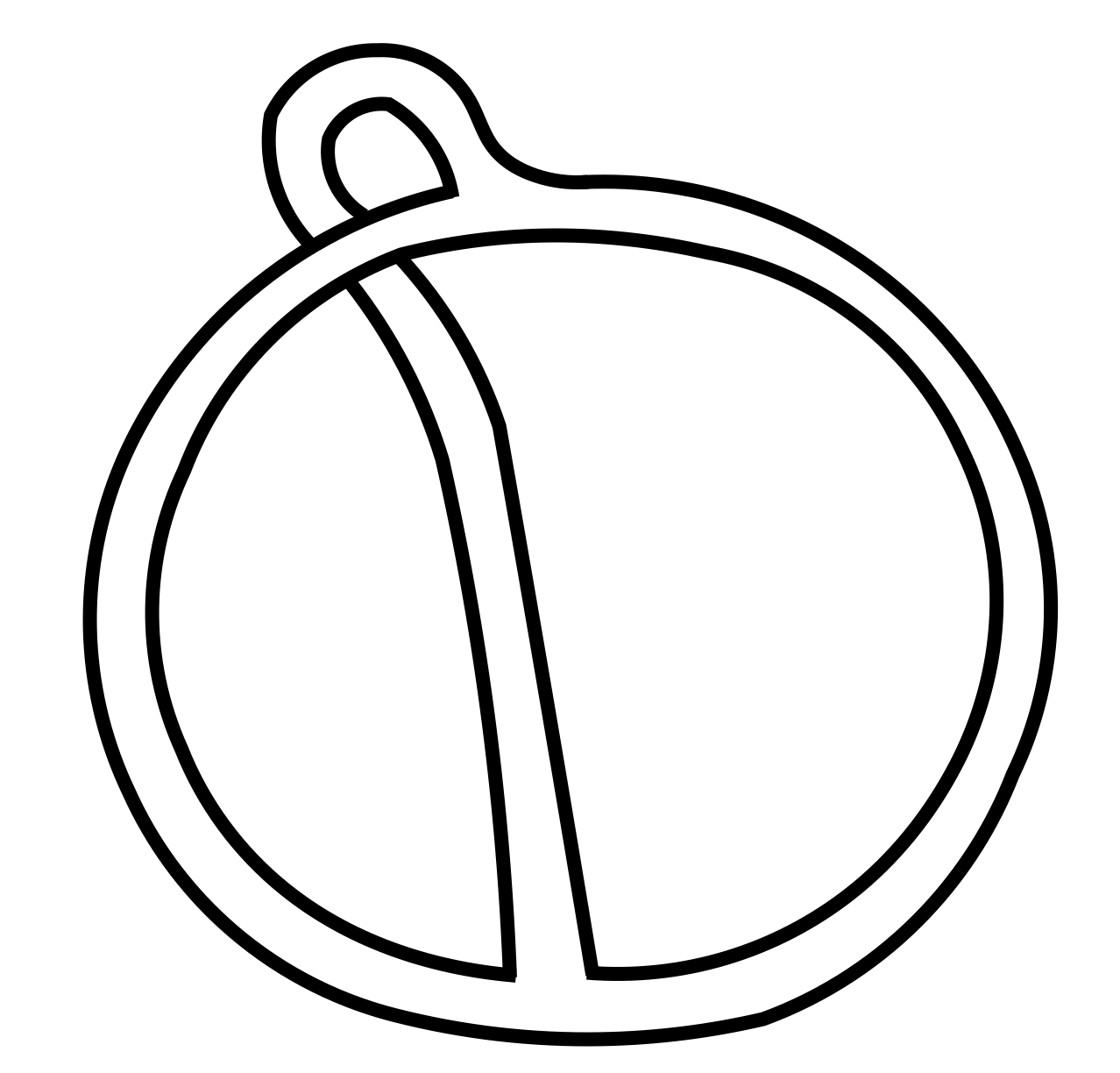}
  \caption{\footnotesize Ribbon $R_{1,1}\in\overline{\mathcal{M}}_{1,1}$}
  \label{fig:ribbon1}
\end{subfigure}
\begin{subfigure}{.2\textwidth}
  \centering
  \includegraphics[width=0.78\textwidth]{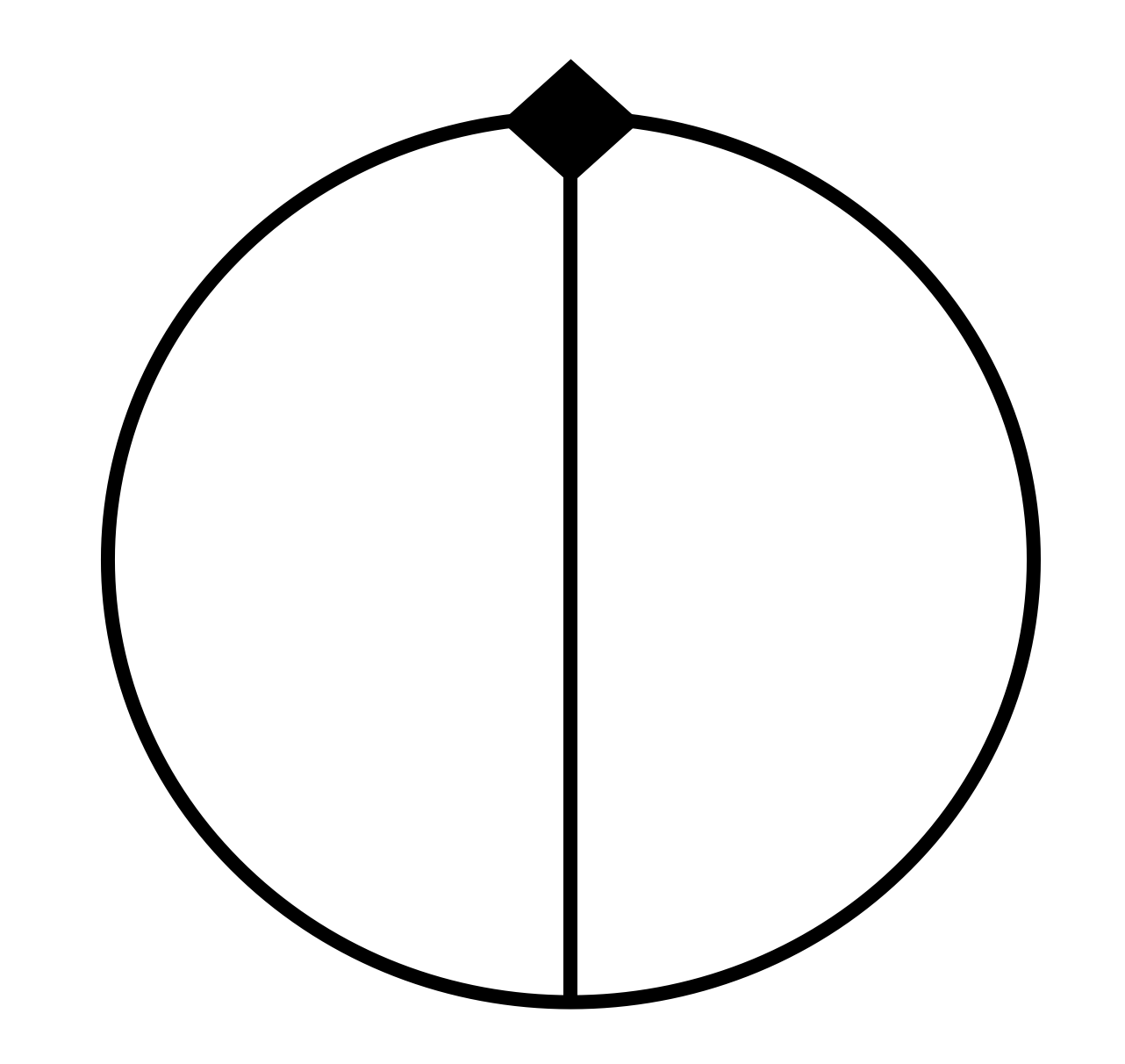}
  \caption{\footnotesize Graph $\Gamma_{1,1}$ for $R_{1,1}$}
  \label{1D-graph-1}
\end{subfigure}
\begin{subfigure}{.45\textwidth}
\hspace{0.7cm}\includegraphics[width=0.9\textwidth]{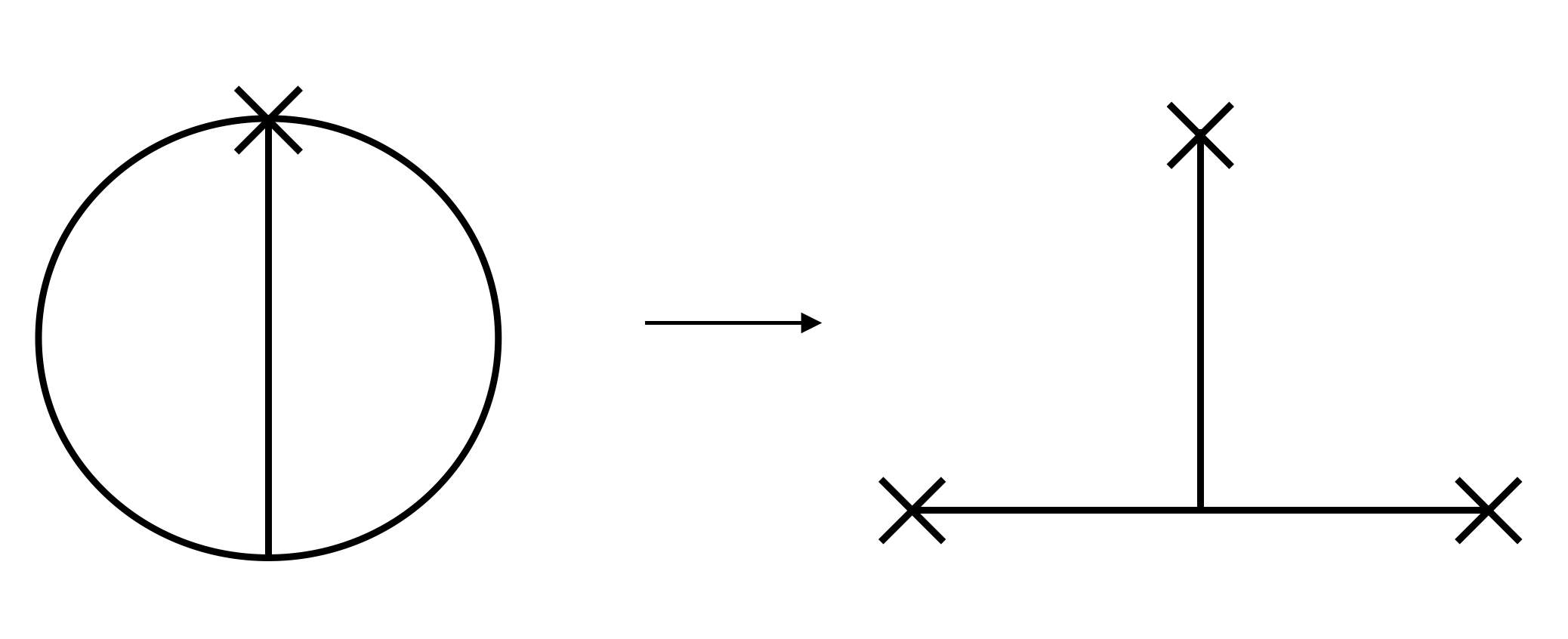}
  \caption{\footnotesize Splitting of a vertex turning $\Gamma_{1,1}$ into $\Gamma_{0,4}$}
  \label{fig:graphx}
\end{subfigure}
\begin{subfigure}{.21\textwidth}
  \includegraphics[width=0.9\textwidth]{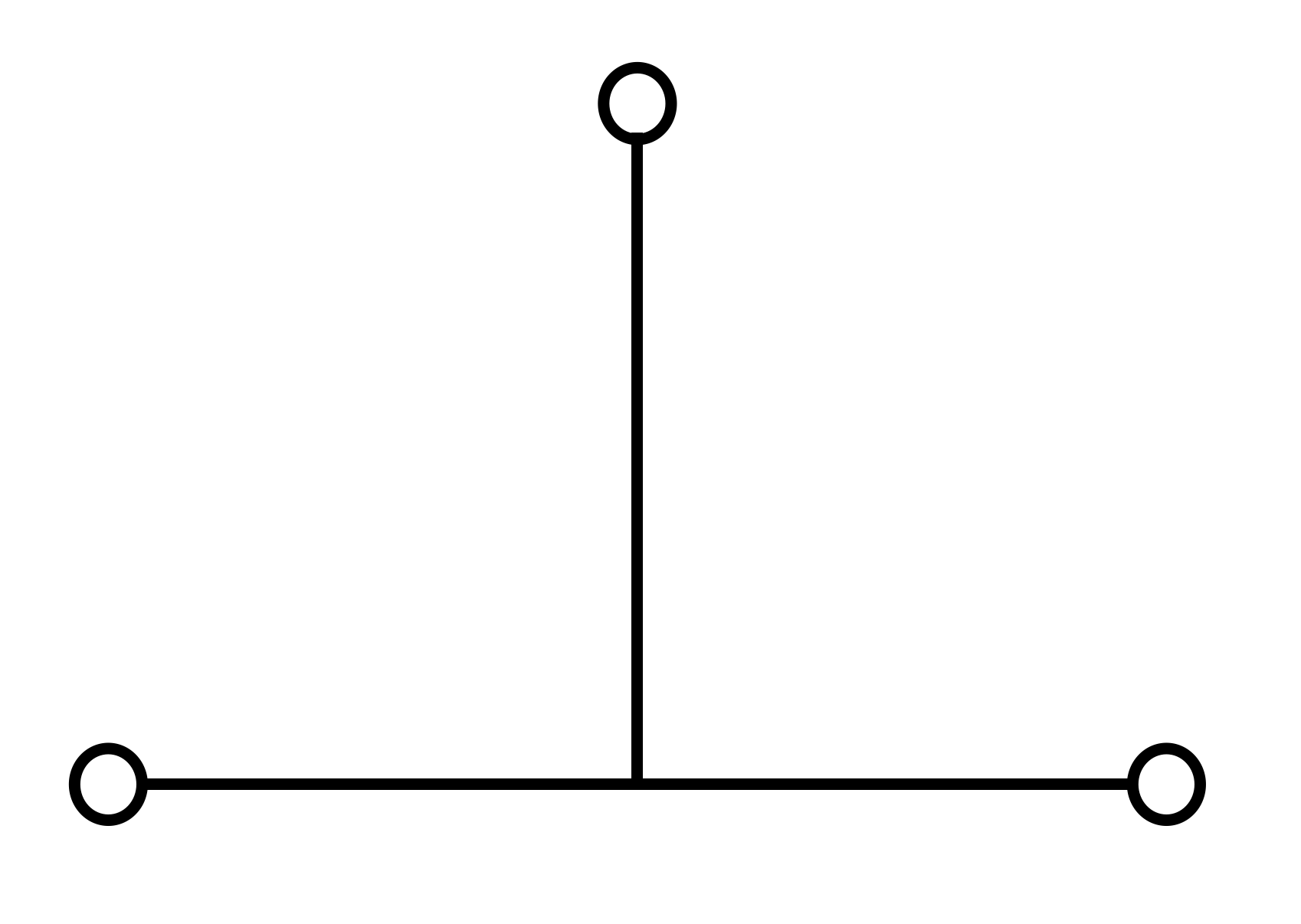}
  \caption{\footnotesize Graph $\Gamma_{0,4}$ for $R_{0,4}$}
  \label{trix}
\end{subfigure}
\hspace{0.1cm}
\begin{subfigure}{.21\textwidth}
  \centering
  \includegraphics[width=0.9\textwidth]{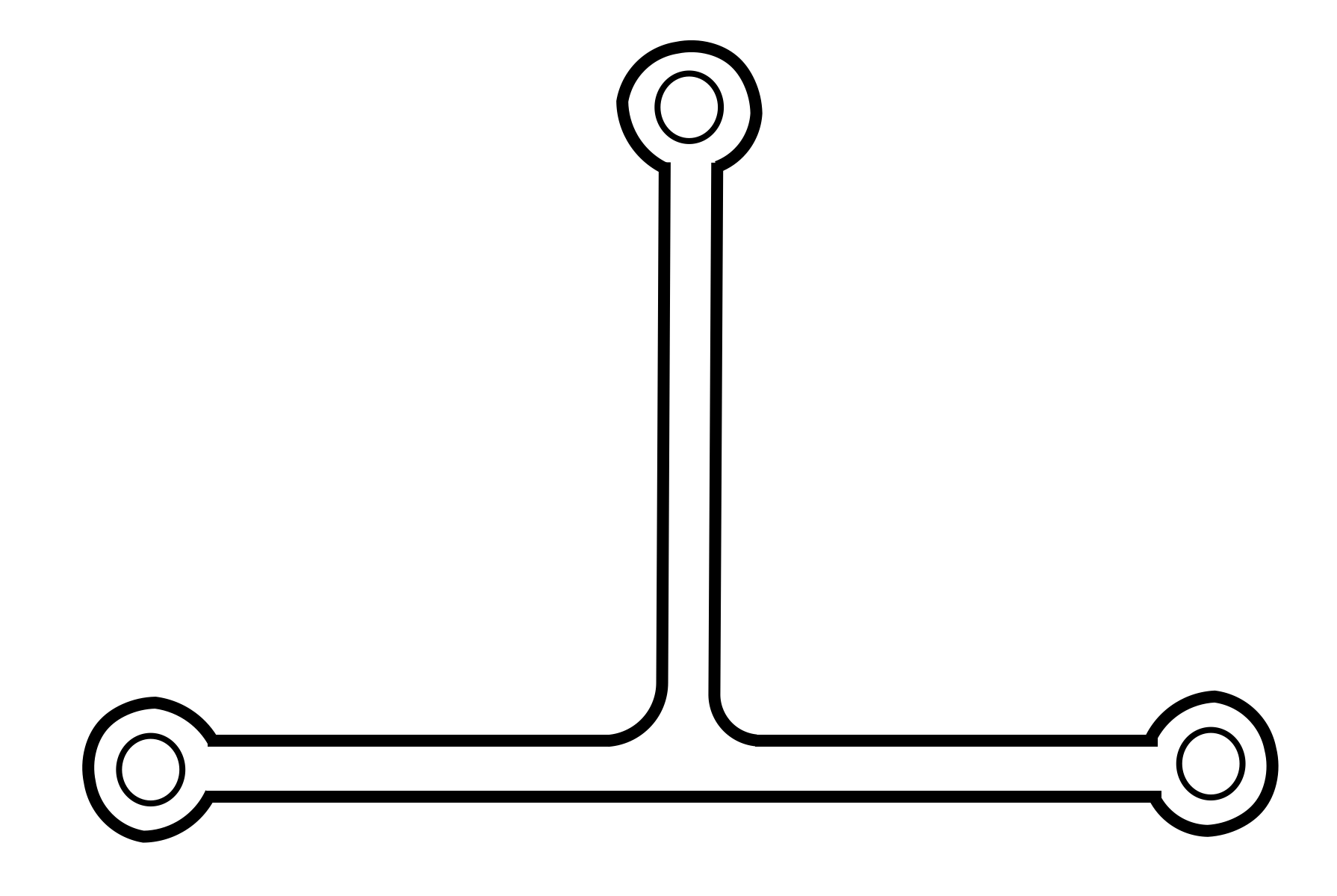}
  \caption{\footnotesize Ribbon $R_{0,4} \in \overline{\mathcal{M}}_{0,4}$}
  \label{thinkt}
\end{subfigure}
    \caption{\small Unique ribbon graph for $\overline{\mathcal{M}}_{1,1}$ and its reinterpretation as a diagram for $\overline{\mathcal{M}}_{0,4}$ with three cusps. }
    \label{1-1-transformation}
\end{figure}

{\bf Effective description---}At every $g$, the leading $\ell=0$ term in \cref{eq:zgl} comes from the highest power of $b$ in \cref{eq:mirz-vol}. 
Similarly, higher-$\ell$ corrections correspond to subleading powers of $b$.
Hence $F_0$ can be obtained from the WP volumes $V_{g,1}(b)$ by taking the boundary size $b$ large. In this limit, hyperbolic geometries become arbitrarily thin strips whose moduli space is described by Kontsevich's ribbon graphs \cite{Kontsevich:1992ti} (see e.g. \cref{fig:ribbon1} and \cref{sec:kont}).

\paragraph{Genus sum as a gas of cusps---} At strictly $b\to\infty$, these ribbons degenerate into one-dimensional trivalent graphs. The resulting vertices are singular points where the original geometry suffers from a three-way pinch-off.
At these points the geometry would thus seem to be splittable at the cost of introducing three cusps, i.e., boundaries of vanishing length. This geometric picture is borne out at the level of the WP volumes: 
the genus cycles of any ribbon graph for $V_{g,1}(b)$ can be fragmented at large $b$ to give rise to a tree graph representing a geometry in $\overline{\mathcal{M}}_{0,1+n}$ where $n=3g$ boundaries are zero-size, and one that remains is of size $b$. We denote the WP volume associated to these cuspy geometries by
\begin{equation}
\label{eq:cuspnot}
    \mathring{V}_n(b) \equiv V_{0,1+n}(b,0,\dots,0).
\end{equation}

The explicit realization of this result relies on two facts. Firstly, the dimensions of moduli spaces $\overline{\mathcal{M}}_{g,1}$ and $\overline{\mathcal{M}}_{0,1+3g}$ are identical, and thus give matching pre-factors to the intersection numbers in \cref{eq:mirz-vol}.
Secondly, the highest-$b$ powers of both WP volumes agree, and the corresponding intersection numbers for the cusp volumes $V_{0,1+n}$ are simply
\begin{equation}
\label{eq:l0sim}
    \expval{\psi^{n-2}}_{0,1+n} = 1.
\end{equation}
As a result, we obtain the large-$b$ identity (cf. \cref{eq:g-int-number})
\begin{equation}
\label{eq:wprel}
\boxed{
    V_{g,1}(b) \simeq \frac{1}{24^gg!} \mathring{V}_{3g}(b),
    }
\end{equation}
with corrections down by $1/b^2$. In light of this key result, we extend \cref{eq:zgb} to define a cusp partition function
\begin{equation}
\label{eq:z0bn}
    Z_0(\beta,0^n) \equiv \int_0^\infty b\, db \,  Z_{\smalltext{T}}(\beta,b) \, \mathring{V}_n(b),
\end{equation}
which in a large-$\beta$ expansion analogous to \cref{eq:zgl} gives
\begin{align}
    Z_0(\beta,0^n) \reprel{\ell=0}{=} \frac{\beta^{n}}{\sqrt{2\pi} \beta^{3/2}}.
\end{align}
For $n=3g$, this is related to $Z_g(\beta)$ just like in \cref{eq:wprel}. Hence the genus sum for $\expval{Z(\beta)} $ at large $\beta\sim e^{2S_0/3}$ can be equivalently interpreted as a grand canonical ensemble of cusp triplets on a single disk topology. Explicitly,
\begin{align}
\label{eq:cusp-sum}
    \expval{Z(\beta)} \reprel{\ell=0}{=} e^{S_0}\sum_{g=0}^\infty \frac{\lambda^g}{g!} Z_0(\beta,0^{3g}), \quad \lambda = \frac{e^{-2S_0}}{24},
\end{align}
with $\lambda$ the fugacity. This allows for a simple physical interpretation of the results of the previous section. Namely, all geometric structure of higher topologies degenerates at large $\beta\sim e^{2S_0/3}$ into a gas of cusp triplets, with each genus replaced by three cusps. 
In the gas, the $1/g!$ factor encodes the fact that genera/cusps carry no labels, and are therefore indistinguishable.
The upshot is a reformulation of the large-$\beta$ sum over all topologies as a gas of cusps in JT on just a fixed disk topology.
\begin{figure}[ht!]
\includegraphics[valign=c,scale=0.25]{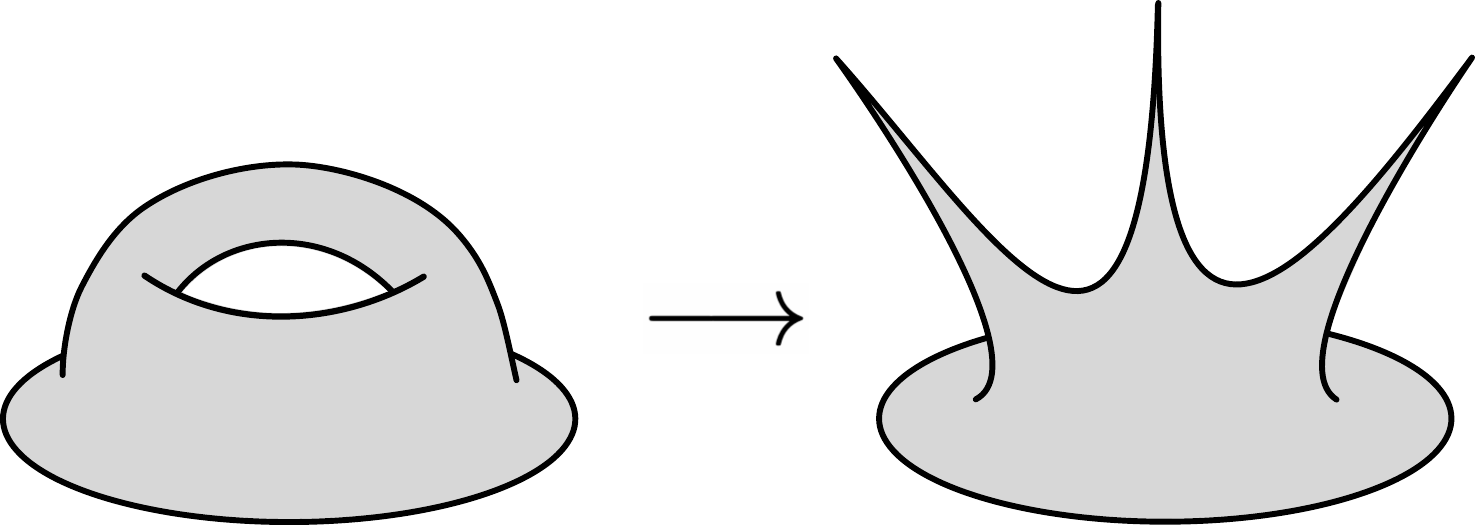}
    \caption{Schematic replacement of genus by cusps.}
\end{figure}

\paragraph{Effective description---}The replacement of genus by cusps grants an alternative geometric description, corresponding to a resummation of the cusp expansion into a particular deformation of the JT dilaton potential.

Recall the JT action on a space $M$ of disk topology,
\begin{align}
    S_{\smalltext{JT}} = - S_0 -\frac{1}{2}\int_M \phi\,(R+2) - \int_{\partial M} \phi\,(K-1).
\end{align}
A conical defect with opening angle $2\pi-\alpha$ can be created on $M$ by inserting in the GPI the operator
\begin{equation}
    Q \equiv q\int_M e^{-\alpha\phi},
\end{equation}
where $\alpha=2\pi$ for cusps.
The constant $q$ is to be fixed by a proper normalization to match the WP volumes.
Letting $M$ be a disk with a single length-$\beta$ asymptotic boundary and inserting $n$ cusps,
\begin{equation}
\label{eq:cuspins}
    e^{S_0} Z_0(\beta,0^{n}) = \int_{\smalltext{Disk}} \mathcal{D} g \mathcal{D} \phi \, e^{-S_{\smalltext{JT}}} \, {Q}^n,
\end{equation}
which realizes the cusp partition function in \cref{eq:z0bn}. Using \cref{eq:cuspins} in \cref{eq:cusp-sum}, the sum over $g$ exponentiates into a deformation of the JT dilaton potential,
\begin{align}
    \expval{Z(\beta)} \reprel{\ell=0}{=} \int_{\smalltext{Disk}} \mathcal{D} g \, \mathcal{D} \phi \, e^{-S}, \quad S\equiv S_{\smalltext{JT}} - \lambda {Q}^{3} \label{eq:deformed-theory}
\end{align}
For a single $Q$, this type of deformation has been studied extensively at different $\alpha$ values \cite{Witten:2020ert, Witten:2020wvy, Turiaci:2020fjj, Lin:2023wac}. For instance, when $\alpha \geq \pi$, the resulting theory can be studied using WP volumes with boundaries of imaginary lengths. The quantization is subtle as the deformed theory has to be appropriately renormalized. In addition, here ${Q}$ appears cubed, thus giving a nonlocal deformation of the theory.

Fortunately, the nonlocality sourced by ${Q}^3$ is sufficiently tame that our theory remains amenable to the same quantization as in \cite{Witten:2020ert}. In particular, a na\"ive classical evaluation of the on-shell action still provides the correct functional answer up to parameter renormalizations which can then be fixed using known results.

One begins with a black hole ansatz on a disk topology of renormalized boundary length $\beta$ of the form
\begin{align}
g = A(r) \, d\tau^2 + A(r)^{-1} dr^2, \quad r\geq r_h, \quad \tau\sim\tau+\beta.
\end{align}
On this geometry and with a radial ansatz $\phi=\phi(r)$, the bulk action evaluates to
\begin{align}
\label{eq:bulkon}
\!\!-S_0 - \frac{\beta}{2} \!\int\! dr \phi  (A''(r) - 2) -\lambda q^3 \beta^3 \left(\int dr\, e^{-\alpha \phi} \right)^3\!\!.
\end{align}
Extremizing with respect to $\phi$ leads to
\begin{align}
\label{eq:phieom}
A''(r) - 2 = -6 \lambda q^3 \beta^2 \alpha e^{-\alpha \phi} \left(\int dr\, e^{-\alpha \phi} \right)^2.
\end{align}
Varying $A(r)$ yields the equation of motion $\phi''(r) = 0$. 
By reparameterizations of $\tau$, $r$, and $A$, the linear solution can be written $\phi(r)=r$ \footnote{More specifically, the metric is invariant under $r\to r/a-b$, $A\to A/a^2$, and $\tau\to a\tau$. Hence diffeomorphism invariance can be used to turn a general solution of the form $\phi=ar+b$ into simply $\phi(r)=r$.}. 
The condition for a smooth horizon, $\beta = 4\pi/A'(r_h)$, gives
\begin{align}
r_h = \frac{2\pi}{\beta} - \frac{3\lambda \beta^2}{\alpha^2} e^{-3\alpha r_h} .
\end{align}
For $\beta\sim e^{2S_0/3}$ both terms on the right are of the same order and make $r_h \sim e^{-2S_0/3}$. Thus at leading order
\begin{align}
S_* = -S_0 -\frac{\lambda q^3 \beta^3}{\alpha^3},
\end{align}
with corrections suppressed by $e^{-2S_0/3}$. Therefore,
\begin{align}
\label{eq:sad}
\expval{Z(\beta)} \simeq e^{S_0} Z_{\smalltext{1-loop}} \exp\left(\frac{q^3 \beta^3 e^{-2S_0}}{24 \alpha^3} \right),
\end{align}
after recalling $\lambda = e^{-2S_0}/24$. The one-loop factor accounts for boundary fluctuations governed by the Schwarzian action. As in standard JT, on the disk we have three zero-modes which lead to $Z_{\smalltext{1-loop}}= 1/\sqrt{2\pi \beta^3}$. This is because evaluated on the disk with no cusp insertions, the isometry group is still $SL(2,\mathbb{R})$.

Given this saddle-point analysis, one must ask how renormalization modifies \cref{eq:sad}. 
The scheme advocated for by \cite{Witten:2020wvy} involves comparison with the sum over cusps in \cref{eq:cusp-sum}. Alternatively, \cite{Turiaci:2020fjj} proposed performing the computation in a general $(2,p)$ minimal model, and then taking the $p\to \infty$ for JT gravity. Both prescriptions agree, and they result in $q=\alpha$ (to leading order as $\alpha\to 2\pi$). Additionally, our answer is the only term contributing to $\expval{Z(\beta)}$ if and only if $\alpha=2\pi$. Altogether, our effective theory on the disk precisely matches the full genus expansion of JT gravity to leading order at large $\beta\sim e^{2S_0/3}$.

Our cusp deformation also induces a shift of the spectral edge of the theory down to a lower energy $E_0<0$ relative to that of JT. As a function of $r_h$, the thermodynamic energy is given by
\begin{equation}
\!\!E(r_h) = \!\!\lim_{r_\infty \to \infty} \frac{1}{2} \left(r_\infty^2 \! - A(r_\infty)\right) = \frac{r_h^2}{2} - 3\lambda \beta^2 e^{-3\alpha r_h}.
\end{equation}
The minimum energy is found to occur at $r_h = 9\lambda \beta^2$, and to leading order is given by
\begin{equation}
\label{eq:e0shift}
E_0 = -3\lambda\beta^2 = -\frac{1}{8} \beta^2 e^{-2S_0}.
\end{equation}
This precisely matches the saddle-point location of the single eigenvalue instanton found by \cite{Hernandez-Cuenca:2024icn} at $\beta\sim e^{S_0}$.
Here though $\beta\sim e^{2S_0/3}$, so $E_0\sim e^{-2S_0/3}$ still remains in the neighborhood of the original JT edge.

\paragraph{Subleading effects and other observables---}Higher-order corrections to \cref{eq:lzero} come from $\ell\geq1$ in \cref{eq:zgl}. The genus sums for additional $F_\ell$ functions can be computed using results from \cite{Okuyama:2019xbv}, and are quoted in \cref{sec:higherl} for small $\ell$. In the effective cusp theory, we expect these corrections to be captured by higher-order interactions among cusps. For a general cusp deformation
\begin{equation}
\label{eq:sqdef}
    S_Q \equiv -\sum_{k=3} \lambda_k Q^{k},
\end{equation}
we already found that $\ell=0$ fixes $\lambda_3=\lambda$. The first correction coming from $\ell=1$ imposes $\lambda_5= \frac{21}{5}\pi^2 \lambda^2$. Higher-$\ell$ corrections continue to introduce additional cusp interactions, as well as renormalize some of the couplings. For instance, $\ell=3$ activates $\lambda_6\sim O(\lambda^3)$ and $\lambda_9\sim O(\lambda^4)$, but also introduces an $O(\lambda^2)$ correction to $\lambda_3$. These results are explained in detail in \cref{sec:subleadcusp}, and require novel expressions for the general intersection numbers that compute cusp WP volumes which we address in \cref{sec:higherlc}.

Given the matching of $\langle Z(\beta)\rangle$ between the genus and cusp expansions of JT, one may ask whether other observables are also equal. This is indeed the case to leading order when computing
\begin{align}
    \langle Z(\beta) Z(\gamma_1)\cdots Z(\gamma_m) \rangle,
\end{align}
with $Z(\gamma_i)$ being probe boundaries with $\gamma_i\ll\beta$ \footnote{One could also try to consider $\langle Z(\beta_1) Z(\beta_2)\rangle$ for $\beta_1$ and $\beta_2$ both nonperturbatively large. However, by inserting a second large boundary we are once again deforming the background of the theory in the same way we did when we inserted one large boundary. This may also lead to an effective theory, but it would be a different one.}. 
This follows from the remarkable fact that intersection numbers on the genus side continue to give
\begin{align}
   \expval{\psi_1^{3g-2+m}}_{g,1+m} = \frac{1}{24^g g!}.
\end{align}
This can be calculated from \eqref{eq:g-int-number} by repeatedly applying the string equation \cite{Witten:1990hr} to remove the $m$ extra punctures.

{\bf Geometric transition from genus to cusps---}Before concluding, we provide a more refined geometric picture for how each genus of a surface in $\overline{\mathcal{M}}_{g,1}$ degenerates into three cusps in the Airy limit where its geodesic boundary length $b$ becomes large. 
Recall that the volume of a general hyperbolic surface in $\overline{\mathcal{M}}_{g,1}$ is a topological invariant given by the Euler characteristic, thus independent of $b$. This means that as $b$ grows, for the volume to remain constant, any surface must narrow down to a thin strip close to the boundary. At large $b$, this gives rise to Kontsevich's decomposition of moduli space into trivalent ribbon graphs \cite{Kontsevich:1992ti} \footnote{See also \cite{Blommaert:2022lbh,Saad:2022kfe} for recent discussions of Kontsevich's ribbon graphs in the context of JT gravity.}.

By the identity between WP volumes in \cref{eq:wprel}, the construction in \cref{1-1-transformation} explains how the moduli space of ribbon graphs in $\overline{\mathcal{M}}_{g,1}$ relates to that of cuspy ones in $\overline{\mathcal{M}}_{0,1+3g}$ at every $g$. A characterization of the moduli subspace of $\overline{\mathcal{M}}_{0,1+n}$ involving cuspy geometries is given in \cref{sec:cuspchar}. Since the combinatorics become very involved as $g$ increases, here we only illustrate the $g=1$ case. 

The idea is as follows: there exists a single type of ribbon geometry in $\overline{\mathcal{M}}_{1,1}$, shown in \cref{fig:ribbon1}. This is represented by a graph in \cref{1D-graph-1}, where a diamond indicates orientation reversal capturing the non-planarity of the ribbon. The graph may be understood as the $b\to\infty$ limit of the ribbon, in which the geometry degenerates and vertices become singular three-way points. These may be interpreted as singular endpoints for each of the three incident edges where the graph can be punctured open. Doing so at e.g., the diamond vertex, the graph splits as in \cref{fig:graphx}. Restoring trivalency at every leaf of the resulting tree graph, one arrives at the ribbon graph in \cref{trix}. The ribbon this represents is shown in \cref{thinkt}, and corresponds to a geometry in $\overline{\mathcal{M}}_{0,4}$ with one large boundary and three infinitesimal ones, i.e., cusps.

This procedure extends similarly to higher-genus ribbon graphs. Namely, a general graph for $\overline{\mathcal{M}}_{g,1}$ involves $E_{g,1} = 6g-3$ edges and $V_{g,1}=4g-2$ vertices (see \cref{eq:kont}), and has multiple cycles. Performing $g$ vertex splittings without disconnecting it leads to a graph with $3g$ degree-$1$ vertices and $3g-2$ degree-$3$ ones. These ingredients correspond precisely to a trivalent tree graph with $3g$ leaves where no cycles remain. Making the leaves degree-$3$ by attaching infinitesimal loop edges, one always arrives this way at a cuspy ribbon graph for $\overline{\mathcal{M}}_{0,1+3g}$.

{\bf Discussion---}Our main result is that the full genus sum of JT gravity at low temperatures can be completely reproduced on a single disk topology by the effective theory in \cref{eq:deformed-theory}. 
This theory involves a highly quantum and nonlocal deformation of the dilaton potential, which we derived using identities between WP volumes that allowed us to rewrite the genus expansion as a gas of cusps.

The example we have described contains elements from both the weak and the strong coupling regimes of a topological expansion. Similarly to the weak coupling story, we have a nonlocal theory on spacetime (or from a string theory perspective, on the worldsheet). This is natural, since each ``wormhole'' corresponding to a genus on the worldsheet interpolates between two disconnected regions. The cost of reducing all the topologies into a disk is working with a nonlocal theory \footnote{For some non-topological perspectives on the issue of locality in quantum gravity, see e.g., \cite{Amelino-Camelia:2011lvm, Almheiri:2014lwa, Donnelly:2016auv, Valdes-Meller:2021com, Hernandez-Cuenca:2024pey}}.

The strong coupling side of JT involves $\beta \sim e^{S_0}$, which gives rise to the single-eigenvalue instanton studied by \cite{Okuyama:2021eju, Hernandez-Cuenca:2024icn}.
Our results for $\beta \sim e^{2S_0/3}$ thus correspond to a small-$\beta$ limit of the instanton regime. This instanton can be understood as arising from open-closed duality in the topological string perspective of JT \cite{McNamara:2020uza,Post:2022dfi}, which in this context is similar to the previously mentioned S-duality relating strings and D1-branes. So the $\beta\sim e^{2S_0/3}$ Airy limit can be thought of as either a strong coupling limit of JT, or a weak coupling limit of an eigenvalue instanton in JT. 
In this sense, the Airy regime can be understood as a critical point of $\expval{Z(\beta)}$ at large $\beta$.

There remain multiple directions worth exploring in the future. 
Inspired by the matrix saddle-point treatment of the eigenvalue instanton in \cite{Hernandez-Cuenca:2024icn}, an original motivation for this work was also to obtain a geometric description of the strong coupling limit. While this remains work in progress, our results here already offer a tantalizing geometric picture. In particular, the instanton is manifestly related to a determinant which on the gravity side corresponds to a brane operator sourcing new dynamical boundary conditions for the GPI. It is thus tempting to think of our cusp description at intermediate coupling as the onset of such new boundaries starting with zero size.
This is also nicely consistent with the finding in \cref{eq:e0shift} that our deformed cusp theory already captures the initial dynamics of the eigenvalue instanton as it separates from the cut.

It would also be interesting to understand the subleading corrections analyzed in \cref{sec:subleadcusp} more from the perspective of the dual double-scaled matrix model. As we saw, away from the Airy limit, cusps begin interacting in groups of more than just triplets. In particular, as shown at the level of WP volumes in \cref{eq:genhk}, there appear linear combinations over different numbers of cusps. Intuitively, this is because to build JT gravity starting from the Airy matrix model (which only has a cubic interaction), one has to include matrix couplings to all orders \cite{Johnson:2020heh}. Diagrammatically these generate non-regular graphs with vertices of higher valency, high upon splitting can generate more arbitrary numbers of cusps.

It is known that JT gravity has a worldsheet description as the large-$p$ limit of $(2,p)$-minimal string theories \cite{Saad:2019lba,Mertens:2020hbs}. Deformations of this theory by tachyon operators $\tau_n$ were studied in \cite{Turiaci:2020fjj}, where it was shown that the defect operator $Q$ should be identified with $\tau_n$ with $n = \frac{p}{4\pi}\alpha$ in the large-$p$ limit (with cusps corresponding to $n = \frac{p}{2}$). Our results suggest that the effective theory with cusps can be equivalently studied as a deformation of the minimal string by an infinite set of higher-point interactions of tachyon operators. It would be interesting to study this theory using CFT techniques as in \cite{Belavin_2009}. Similar statements could be made about the Virasoro minimal string, although the correspondence between the tachyon operators and JT defect operators is not fully understood.

It would also be interesting to see if there are other systems which display an analogous resummation at intermediate coupling. There are $p$-deformed volumes defined for the $(2,p)$-minimal string \cite{Mertens:2020hbs}, and quantum volumes used in the Virasoro minimal string \cite{Collier:2023cyw}, which could display similar relations at finite $p$ to the ones we have used here. Additionally, there are other variants of JT such as JT supergravity, or JT coupled to matter, which could be addressed using similar principles.  Furthermore, one could explore if the late-time spectral form factor can also be described by a single effective geometry \cite{Saad:2022kfe, Blommaert:2022lbh}. 

Lastly, another possibility is to examine topological string theories on various Calabi-Yau manifolds. From the perspective advocated for in \cite{Post:2022dfi, McNamara:2020uza}, the result we have found in JT can be interpreted as a special case of such theories. Hence using e.g., tools from \cite{Bershadsky:1993cx}, a more general picture of intermediate coupling resummation could potentially emerge when probing extreme limits (such as low temperatures) in these theories.

\paragraph{Acknowledgments.}
\begin{acknowledgments}
We would like to thank Elba Alonso-Monsalve, Tom Hartman, Luca Iliesiu, Jon Sorce, and Douglas Stanford for helpful discussions. We are especially grateful to Scott Collier, David Kolchmeyer, and Adam Levine for many interesting and insightful comments.
SHC is supported by the Templeton Foundation via the Black Hole Initiative (award number 62286) and the MIT Center for Theoretical Physics.
NVM is supported by the Hertz Foundation and the MIT Dean of Science Fellowship. WWW is supported by the DOE through QuantISED grant DE-SC0020360. This work was done in part during the workshop ``Holographic Duality and Models of Quantum Computation'' held at Tsinghua Southeast Asia Center in Bali, Indonesia (2024).

The opinions expressed in this publication are those of the authors and do not necessarily reflect the views of the John Templeton Foundation.
\end{acknowledgments}

\appendix
\setcounter{secnumdepth}{1}

\section{Higher-order genus sums}
\label{sec:higherl}

For $n=1$ boundary, the WP volumes from \cref{eq:mirz-vol} read
\begin{equation}
\label{eq:genwp}
    V_{g,1}(b) = \sum_{\ell=0}^{3g-2} \frac{\expval{\omega^{\ell} \psi^{3g-2-\ell}}_{g,1}}{2^{3g-2-\ell} (3g-2-\ell)! \ell!} b^{2({3g-2-\ell})}.
\end{equation}
The intersection numbers above can be written as
\begin{equation}
\label{eq:intg}
    \expval{\omega^{\ell}\psi^{3g-2-\ell}}_{g,1} = \frac{(2\pi^2)^\ell}{24^gg!} P_{\ell}(g),
\end{equation}
where $P_{\ell}(g)$ is a $2\ell$-degree polynomial in $g$. From \cite{Okuyama:2019xbv},
\begin{equation}
\begin{aligned}
\label{eq:pielgis}
    P_0(g) &= 1, \\
    P_1(g) &= 1 + \frac{12}{5}g(g-1), \\
    P_2(g) & = \frac{1}{175} (g\!-\!1)(1008g^3\!-\!1200g^2\!+\!888g\!-\!175), \\
    P_3(g) & = \frac{1}{875}(g\!-\!1)(12096g^5\!-\!31104g^4\\&
    \qquad\qquad\!+\!35856g^3\!-\!25644g^2\!+\!7960g\!-\!875),
\end{aligned}
\end{equation}
etc. Hence the genus resummations in \cref{eq:lexp} all read
\begin{equation}
    F_\ell(x) = (2\pi^2)^\ell e^{x/24} f_\ell(x),
\end{equation}
where $f_\ell(x)$ is a $2\ell$-degree polynomial in $x$. By \cref{eq:pielgis},
\begin{equation}
\begin{aligned}
\label{eq:felxs}
    f_0(x) &= 1, \\
    f_1(x) &= 1 + \frac{1}{240} x^2, \\
    f_2(x) & = 1 - \frac{1}{24}x + \frac{1}{40}x^2 + \frac{1}{630}x^3 + \frac{1}{57600}x^{4}, \\
    f_3(x) & = 1-\frac{1}{24}x+\frac{4}{45} x^2+\frac{1163}{40320} x^3 \\&
    \qquad+\frac{13}{8960} x^4+\frac{1}{50400}x^5+\frac{1}{13824000}x^6.
\end{aligned}
\end{equation}
Higher-$\ell$ corrections to \cref{eq:lexp} can be computed similarly.

\section{New cusp Weil-Petersson volumes}
\label{sec:higherlc}

The genus $g=0$ cusp WP volumes in \cref{eq:cuspnot} read
\begin{equation}
\label{eq:cuspyint}
    \mathring{V}_n(b) = \sum_{\ell=0}^{n-2}  \frac{\expval{\omega^{\ell} \psi^{n-2-\ell}}_{0,1+n}}{2^{n-2-\ell} (n-2-\ell)! \ell!} b^{2(n-2-\ell)},
\end{equation}
where we used \cref{eq:mirz-vol}. To our knowledge, there do not exist general expressions for the intersection numbers above at fixed $n$ and arbitrary $\ell$ analogous to \cref{eq:intg,eq:pielgis}. Here we derive them using the recursion \cite{do2009weil}
\begin{equation}
    \!\!\!V_{g,1+n}(b_1,\dots,b_n,2\pi i) \!=\! \sum_{i=1}^n \!\int\! db_i b_i  V_{g,n}(b_1,\dots,b_n).\!
\end{equation}
For $g=0$, this relation together with the $\mathbb{S}_{n+1}$ permutation symmetry of $V_{g,1+n}$ allows to uniquely determine the full WP volume $V_{0,1+n}$ solely from $V_{0,n}$ \footnote{Cf. Mirzakhani’s general recursion relation \cite{Mirzakhani:2006eta}, which for $g=0$ requires not only $V_{0,n-1}$ but also the pairs $V_{0,n_1}$ and $V_{0,n_2}$ for all $n_1+n_2 = n+1$.}. In this way, one can obtain explicit formulae for the intersection numbers in \cref{eq:cuspyint}. Beyond the $\ell=0$ case in \cref{eq:l0sim}, for general $\ell\geq1$ it is convenient to write
\begin{equation}
    \expval{\omega^{\ell} \psi^{n-2-\ell}}_{0,1+n} = \frac{\pi^{2\ell} (n-1)!}{(n-2-\ell)!} \, Q_\ell(n),
\end{equation}
where $Q_\ell(n)$ is a degree-$(\ell-1)$ polynomial in $n$. By direct evaluation, we found the following explicit expressions:
\begin{equation}
\label{eq:qielgis}
\begin{aligned}
    Q_1(n) &= 1, \\
    Q_2(n) &= \frac{1}{3}(3n-2), \\
    Q_3(n) &= \frac{1}{3}(3n^2-3n+1), \\
    Q_4(n) &= \frac{1}{15}(15n^3-15n^2+10n-2), \\
    Q_5(n) &= \frac{1}{45}(45n^4 - 30n^3 +45n^2 - 10n+2).
\end{aligned}
\end{equation}
We attach results for $Q_\ell$ up to $\ell=28$ in an ancillary file.

\section{Subleading cusp interactions}
\label{sec:subleadcusp}

Higher-$\ell$ corrections in \cref{eq:lexp} at large $\beta\sim e^{2S_0/3}$ come from perturbative interactions in the effective cusp theory. These can be worked out by reconstructing the genus WP volumes $V_{g,1}(b)$ in \cref{eq:genwp} in terms of the cusp WP volumes $\mathring{V}_n(b)$ in \cref{eq:cuspyint} order by order in $b$.
At leading order $\ell=0$, by \cref{eq:wprel} we already have
\begin{equation}
    24^g g!\,V_{g,1}^{(0)}(b) = \mathring{V}^{(0)}_{3g}(b).
\end{equation}
Lifting this relation to $\ell=1$ using $P_1(g)$ from \cref{eq:pielgis} and $Q_1(n)$ from \cref{eq:qielgis}, there appears an $O(b^{2(3g-3)})$ correction. This is naturally reproduced by the cusp WP volume $\mathring{V}_{3g-1}(b)$ at order $\ell=0$, namely,
\begin{equation}
    24^gg! \, V^{(1)}_{g,1}(b) = \mathring{V}^{(1)}_{3g}(b) - \frac{21}{10} (2\pi^2) g(g-1) \mathring{V}^{(0)}_{3g-1}(b).
\end{equation}
Going to order $\ell=2$ on the left now activates both $\mathring{V}^{(2)}_{3g}$ and $\mathring{V}^{(1)}_{3g-1}$ on the right. Using the results from \cref{sec:higherl,sec:higherlc} for $\ell=2$, one now finds that an $\ell=0$ contribution from $\mathring{V}_{3g-2}$ must be included. In particular, one must add
\begin{equation}
    \frac{3}{1400} (2\pi^2)^2 g(g\!-\!1) (1029 g^2\!-\!3175 g\!+\!2024)
    \mathring{V}^{(0)}_{3g-2}(b),
\end{equation}
to reconstruct $V_{g,1}$ up to $\ell=2$. This strategy can be iterated for up to arbitrary $\ell$, leading to a general relation
\begin{equation}
\label{eq:genhk}
    24^gg! \, V^{(\ell)}_{g,1}(b) = \sum_{k=0}^{\ell} (2\pi^2)^k\,H_k(g)\mathring{V}^{(\ell-k)}_{3g-k}(b),
\end{equation}
where importantly $H_k(g)$ does not depend on $\ell$, and one consistently finds that $H_k(g)=0$ for $3g<k$. The results above can be summarized and extended as follows:
\begin{equation}
\label{eq:hks}
\begin{aligned}
    H_0(g) &= 1, \\
    H_1(g) &= - \frac{21}{10} g(g-1), \\
    H_2(g) &= \frac{3}{1400} g(g-1) (1029 g^2-3175 g+2024), \\
    H_3(g) &= -\frac{1}{42000} g (g-1)(64827 g^4-535248 g^3 \\&
    \qquad~~+1532682 g^2-1769853 g+688030).
\end{aligned}
\end{equation}
Extending \cref{eq:genhk} to all orders and integrating against the trumpet in \cref{eq:tpet},
\begin{equation}
    Z_g(\beta) = \frac{1}{24^gg!} \sum_{\ell=0}^{3g} (2\pi^2)^\ell\,H_\ell(g) Z_0(\beta,0^{3g-\ell}),
\end{equation}
thus relating $Z_g$ at a given $\ell$ order to an appropriate combination of cusp partition functions. Recalling \cref{eq:cuspins} and using $\expval{\,\cdot\,}_{\smalltext{JT}}$ as a shorthand for the GPI on the disk,
\begin{equation}
    Z_g(\beta) = \frac{1}{24^gg!} \sum_{\ell=0}^{3g} (2\pi^2)^\ell\,H_\ell(g) \expval{{Q}^{3g-\ell}}_{\!\!\smalltext{JT}}\!\!\!.
\end{equation}
Note that under the GPI one formally has ${Q} \sim \beta$ and the combination $\lambda Q^3_{2\pi} \sim x$ from \cref{eq:lexp}.
Locally replacing $\lambda Q^3_{2\pi}$ by $x$ and resumming over genus leads to
\begin{equation}
\label{eq:glexp}
    \!\!\!\langle Z(\beta) \rangle \!=\! \expval{\!e^{S_0+\lambda {Q}^3} 
    \!\!\left(\!1\!+\!
    \sum_{\ell=1}^{\infty}
    \left(
    \frac{2\pi^2}{{Q}}
    \right)^\ell
    \!\!x^{\left\lceil\frac{\ell+1}{2}\right\rceil} \, \!G_\ell(x)\!
    \right)\!}_{\!\!\smalltext{JT}}\!\!\!\!,
\end{equation}
where $G_\ell(x)$ is a degree-$\ell$ polynomial in $x$,
\begin{equation}
\begin{aligned}
    G_1(x) &= - \frac{21}{10}x,\\
    G_2(x) &= \frac{3}{1400} (1029 x^2+1970 x-210),\\
    G_3(x) &= -\frac{1}{42000} (64827 x^4+372330 x^3\\&
    \qquad\qquad+280935 x^2-62400 x+34300).
\end{aligned}
\end{equation}
At large $\beta \sim e^{2S_0/3}$ we have $x \sim O(1)$, and can expand at large $1/{Q} \sim 1/\beta$ (cf. \cref{eq:glexp}). Taking the logarithm, expanding this way, and then regrouping terms by powers of ${Q}$ we finally obtain the desired cusp couplings. 
Restoring $\lambda {Q}^3$ for $x$, the only interaction at $O(1/\beta)$ is
\begin{equation}
    \frac{S^{(1)}}{2\pi^2} = - \frac{21}{10} \lambda^2 {Q}^5,
\end{equation}
and is fully fixed by $\ell=1$.
At $O(1/\beta^2)$ we get
\begin{equation}
    \frac{S^{(2)}}{(2\pi^2)^2} =
    -\frac{9}{20}\lambda ^2 {Q}^4
    +
    \frac{591}{140}\lambda ^3 {Q}^7,
\end{equation}
which is determined by both $\ell=1,2$. At $O(1/\beta^3)$,
\begin{equation}
    \frac{S^{(3)}}{(2\pi^2)^3} =-\frac{49}{60}\lambda ^2 {Q}^3+\frac{52}{35}\lambda ^3 {Q}^6-\frac{855}{112} \lambda ^4 {Q}^9,
\end{equation}
fixed by $\ell\leq3$, and so on. Notice how the latter renormalizes the coupling of the original ${Q}^3$ interaction of the effective cusp theory. For a general cusp deformation written as in \cref{eq:sqdef}, our findings using up to $\ell=3$ are
\begin{equation}
\begin{aligned}
    \lambda_3 &= \lambda + (2\pi^2)^3 \frac{49}{60} \lambda^2 + O(\lambda^3), \\
    \lambda_4 &= (2\pi^2)^2\frac{9}{20}\lambda ^2 + O(\lambda^3), \\
    \lambda_5 &= (2\pi^2)\frac{21}{10}\lambda ^2 + O(\lambda^3), \\
    \lambda_6 &= -(2\pi^2)^3\frac{52}{35}\lambda^3 + O(\lambda^4), \\
    \lambda_7 &= -(2\pi^2)^2\frac{591}{140}\lambda ^3 + O(\lambda^4), \\
    \lambda_8 &= O(\lambda^4), \\
    \lambda_9 &= (2\pi^2)^3\frac{855}{112}\lambda ^4 + O(\lambda^5).
\end{aligned}
\end{equation}

\section{Kontsevich's ribbon graphs}
\label{sec:kont}

The WP volumes $V_{g,n}$ for large geodesic boundary lengths reduce to the moduli space volumes of the topological Airy model $V_{g,n}^{\smalltext{Airy}}$. These can be computed by Kontsevich's formula \cite{Kontsevich:1992ti}
\begin{equation}
\label{eq:kont}
\begin{aligned}
    V_{g,n}^{\smalltext{Airy}}&(b_1,\dots,b_n) = \sum_{\Gamma\in\Gamma_{g,n}} V_{\Gamma}(b_1,\dots,b_n), \\
    V_{\Gamma}(b) &\!\equiv \!\frac{2^{2g-2+n}}{\abs{\Aut(\Gamma)}} \!\prod_{k=1}^{E_{g,n}} \!\int_{0}^\infty\!\!\!\!\! d\ell_k \!\prod_{i=1}^n \delta\!\left(b_i - \sum_{k=1}^{E_{g,n}} n_{k}^i \ell_k \right)\!\!,
\end{aligned}
\end{equation}
where the sum is over $\Gamma_{g,n}$, the set of all topologically inequivalent ribbon graphs corresponding to Riemann surfaces of genus $g$ and $n$ boundaries. Each such $\Gamma\in\Gamma_{g,n}$ consists of $E_{g,n} = 6g-6+3n$ edges and $V_{g,n} = 2g-2+n$ trivalent vertices. The symmetry factor $\abs{\Aut(\Gamma)}$ refers to the cardinality of the group of half-edge automorphisms \cite{Kontsevich:1992ti,normando}. Each variable $\ell_k$ is the length of edge $k$, and $n_{k}^i\in\{0,1,2\}$ is the number of ribbon sides of edge $k$ that are part of the $i^{\text{th}}$ boundary.

The unique ribbon graph for $\overline{\mathcal{M}}_{1,1}$ is shown in \cref{1D-graph-1} and encodes the geometry in \cref{fig:ribbon1}. Its three edges are each traversed twice in following the full ribbon graph boundary, thus giving its geodesic length as
\begin{equation}
\label{eq:b123}
    b = 2(\ell_1+\ell_2+\ell_3).
\end{equation}
For this graph $\abs{\Aut(\Gamma)} = 6$, and applying \cref{eq:kont} gives
\begin{equation}
\label{eq:11b}
    V_{1,1}^{\smalltext{Airy}}(b) = \frac{2}{6} \!\int_{0}^\infty \!\!\!\!\!\! d\ell_1 d\ell_2 d\ell_3 \, \delta(b - 2(\ell_1+\ell_2+\ell_3)) = \frac{b^2}{48},
\end{equation}
which indeed matches the large-$b$ part of the WP volume
\begin{equation}
    V_{1,1}(b) = \frac{1}{24} \left(\frac{b^2}{2} + 2\pi^2\right).
\end{equation}

For $\overline{\mathcal{M}}_{0,4}$ there are multiple ribbon graphs, but only one contributes when three of the boundaries are shrunk to zero size. The surviving graph is the one in \cref{trix}, which captures the geometry in \cref{thinkt}. The length of each infinitesimal boundary is accounted for by a single edge whose length is forced to zero in the cusp limit. The length of the only nontrivial boundary is given by an expression like \cref{eq:b123} in terms of the remaining edges. The only integral to be performed in \cref{eq:kont} is thus identical to that in \cref{eq:11b}. However, combinatorial factors in this case are different: there are $3!$ inequivalent labelings of the cusp boundaries giving $6$ distinct graphs, and the symmetry is reduced to $\abs{\Aut(\Gamma)}=3$. Altogether,
\begin{equation}
\label{eq:04b}
    \mathring{V}_{3}^{\smalltext{Airy}}(b) = 6\times\frac{2^2}{3} \times \frac{b^2}{16} = \frac{b^2}{2},
\end{equation}
which also matches the large-$b$ part of the WP volume
\begin{equation}
    V_{0,4}(b,b_1,b_2,b_3) = \frac{1}{2}\left(b^2 + b_1^2 + b_2^2 + b_3^2 + 4\pi^2 \right).
\end{equation}

More generally, for $\overline{\mathcal{M}}_{g,1}$ ribbon graphs, the total geodesic length always involves traversing every edge graph twice. Since there are $E_{g,1}=6g-3$ edges, this gives the constraint
\begin{equation}
    b = 2\sum_{k=1}^{6g-3}\ell_k.
\end{equation}
In general, the relevant integral that arises in \cref{eq:kont} is
\begin{equation}
\label{eq:cmn}
    C_{m}(b) \equiv \prod_{k=1}^m \int_0^{\infty} \!\!\!\!\!\!d\ell_k \, \delta\!\left(\! b - 2\sum_{k=1}^{m}\ell_k \!\right) \!= \frac{b^{m-1}}{2^m (m-1)!}.
\end{equation}
In particular, for any $\Gamma \in \Gamma_{g,1}$ graph $m=6g-3$, giving
\begin{equation}
    V_{\Gamma}(b) = \frac{2^{2g-1}}{\abs{\Aut(\Gamma)}} C_{6g-3}(b).
\end{equation}
The total WP volume is thus given by
\begin{equation}
\label{eq:vg1ai}
    \!\!V_{g,1}^{\smalltext{Airy}}(b) = \!2^{2g-1} \!N_{g} C_{6g-3}(b), \!\quad\! N_{g} \!\equiv \!\!\!\!\sum_{\Gamma\in\Gamma_{g,1}} \!\! \frac{1}{\abs{\Aut(\Gamma)}}.
\end{equation}
Obtaining the WP volume $V_{g,1}^{\smalltext{Airy}}(b)$ thus reduces to the combinatorial task of enumerating all possible ribbon graphs $\Gamma \in \Gamma_{g,1}$ and working out their symmetry factors.

As for $\overline{\mathcal{M}}_{0,n}$, ribbon graphs generally involve structurally different integrals in Kontsevich's formula depending on how graph edges participate in different boundaries. However, as explained in \cref{sec:cuspchar}, for cusp WP volumes where only one boundary has nonzero length, the situation is much simpler.

\section{Characterization of cusp ribbon graph}
\label{sec:cuspchar}

For cusp WP volumes, Kontsevich's integral in \cref{eq:kont} contains only one $\delta$ function constraint involving a nonzero geodesic length. Since the integral over $\ell_k$ is nonnegative, the other $\delta$ functions constrain every $\ell_k$ variable inside to be strictly zero. However, if more than one such variable appears within the same $\delta$ function, then the support of the integrand becomes of codimension higher than $1$. In other words, the integrand trivializes to a support of measure zero, thus resulting in no contribution to the cusp volume.
More explicitly,
\begin{equation}
    \int_0^{\infty} d\ell \, \delta( -\ell + x ) = \theta(-x),
\end{equation}
which makes any additional integral over $x>0$ vanish.

This implies that, in general, cusp WP volumes can only receive contributions from ribbon graphs where the cusp boundaries involve strictly only one edge. But the only way in which a single edge makes up a full boundary is if it forms a loop, i.e., if it is a self-edge which starts and ends on the same vertex. By trivalency, any such loop can only connect to the rest of the graph via another edge incident on its vertex (cf. a tadpole). 

In the cusp limit, these boundary loops become of zero size, and can be reduced to labelled endpoints (cf. going from \cref{thinkt} to \cref{trix}). This happens for every cusp, yielding a graph with $n$ degree-$1$ cusp vertices. The remaining large boundary surrounds the whole graph, traversing every edge twice when following the ribbon boundary. All other vertices in the graph are trivalent, as usual. Because $g=0$ graphs are planar and we have a single boundary, the final graph can be easily described as a trivalent tree graph, i.e., a graph with no cycles where all vertices are either degree-$3$ (internal) or degree-$1$ (leaves).

The above provides a very simple characterization of all the ribbon graphs that contribute to cusp WP volumes. Namely, given some number of cusps $n$, one begins by enumerating all possible trivalent tree graphs with $n$ leaves. The internal vertices of these have to then be assigned an orientations in order to construct all possible ribbon tree graphs with $n$ cusps, the collection of which we may denote by $T_{n}$. Because cusps are obtained from labelled boundaries but we get the same contribution by exchanging them, any given such graph should be weighted by a factor of $n!$. Each of them will of course also come divided by a combinatorial factor accounting for its half-edge automorphism group. 

Finally, every $\Gamma\in T_n$ ends up involving the same integral in \cref{eq:kont}, corresponding to \cref{eq:cmn} with $m$ the number of edges of nonzero length, i.e., $m=E_{0,1+n}-n = 2n-3$. Hence for any $\Gamma\in T_n$,
\begin{equation}
    V_{\Gamma}(b) = \frac{2^{n-1}}{\abs{\Aut(\Gamma)}} C_{2n-3}(b),
\end{equation}
and thus the complete cusp WP volume is
\begin{equation}
\label{eq:cusptree}
    \!\!\mathring{V}^{\smalltext{Airy}}_n(b) = 2^{n-1} K_n C_{2n-3}(b), \quad K_n \equiv \!\!\sum_{\Gamma\in T_n}\!\! \frac{n!}{\abs{\Aut(\Gamma)}}.
\end{equation}

For example, \cref{eq:wprel} relates $g=2$ to a cusp WP volume with $n=6$ cusps. There only exist two trivalent tree graphs with $6$ leaves. Turning them into ribbon graphs, one easily verifies that one of them admits two inequivalent vertex orientations with $\abs{\Aut(\Gamma)}=1$ both, and the other is unique with $\abs{\Aut(\Gamma)}=3$.
As advocated by the geometric picture in the main text, all of these are obtainable from splittings of trivalent vertices of $g=2$ ribbon graphs. Finally, one verifies that they add up to the right cusp WP volume,
\begin{equation}
    \mathring{V}^{\smalltext{Airy}}_{6}(b) = 
    \frac{6! b^{8}}{2^4 8!} 
    \left(
    1+1+\frac{1}{3}
    \right) = \frac{b^8}{384}.
\end{equation}

\bibliography{refs}

\end{document}